\documentclass[preprint,preprintnumbers,amsmath,amssymb,showkeys,nofootinbib]{revtex4}
\usepackage{txfonts}
\usepackage{amsfonts}
\usepackage{amsmath}
\usepackage{graphicx}
\usepackage{url}
\usepackage{rotating}

\begin{document}

\title{Large cities are less green}

\author{Erneson A. Oliveira$^{1}$, Jos\'e S. Andrade Jr.$^{1}$, Hern\'an A.
Makse$^{1,2}\footnote{Correspondence to: hmakse@lev.ccny.cuny.edu}$}

\affiliation{ $^1$ Departamento de F\'isica, Universidade Federal do Cear\'a,
60451-970 Fortaleza, Cear\'a, Brasil\\ $^2$ Levich Institute and Physics
Department, City College of New York, New York, New York 10031, USA}

\date{\today}

\begin{abstract}
{\bf We study how urban quality evolves as a result of carbon dioxide emissions
as urban agglomerations grow. We employ a bottom-up approach combining two
unprecedented microscopic data on population and carbon dioxide emissions in the
continental US. We first aggregate settlements that are close to each other into
cities using the City Clustering Algorithm (CCA) defining cities beyond the
administrative boundaries. Then, we use data on $\rm{CO}_2$ emissions at a fine
geographic scale to determine the total emissions of each city. We find a
superlinear scaling behavior, expressed by a power-law, between $\rm{CO}_2$
emissions and city population with average allometric exponent $\beta = 1.46$
across all cities in the US. This result suggests that the high productivity of
large cities is done at the expense of a proportionally larger amount of
emissions compared to small cities. Furthermore, our results are substantially
different from those obtained by the standard administrative definition of
cities, \emph{i.e.} Metropolitan Statistical Area (MSA). Specifically, MSAs
display isometric scaling emissions and we argue that this discrepancy is due to
the overestimation of MSA areas. The results suggest that allometric studies
based on administrative boundaries to define cities may suffer from endogeneity
bias.}
\end{abstract}
\maketitle

Allometry was originally introduced in the context of evolutionary theory
\cite{huxley1936} to describe the correlation between relative dimensions of
parts of body size, for instance brain size in mammals, with changes in overall
body size. In a classical result, Kleiber showed that metabolic rate, $\rm{Y}$,
and the body mass, $\rm{X}$, of a large range of mammal's are related by an
allometric power-law $\rm{Y} = \rm{A} \rm{X}^\beta$, where $\beta = 3/4$ is the
allometric exponent and $\rm{A}$ is a constant \cite{kleiber1961}.

In analogy with biological systems, Bettencourt \emph{et al.}
\cite{bettencourt2007} showed that cities across US obey allometric relations
with population size. Indeed, a large class of human activities can be grouped
into three categories according to the value of the allometric exponent: (a)
Isometric behavior (linear, non-allometric or extensive, $\beta = 1$) typically
reflects the scaling with population size of individual human needs, like the
number of jobs, houses, and water consumption. (b) Allometric sublinear behavior
(hipoallometric, non-extensive, $\beta < 1$) implies an economy of scale in the
quantity of interest because its {\it per capita} measurement decreases with
population size. Hipoallometry is found, for example, in the number of gasoline
stations, length of electrical cables, and road surfaces (material and
infrastructure). (c) Superlinear behavior (hyperallometric, non-extensive,
$\beta > 1$) emerges whenever the pattern of social activity has significant
influence in the urban indicator. Wages, income, growth domestic product, bank
deposits, as well as rates of invention measured by patents and employment in
creative sectors, display a superlinear increase with population size. These
superlinear scaling laws indicate that larger cities are associated with optimal
levels of human productivity and quality of life; doubling the city size leads
to a larger-than-double increment in productivity and life standards
\cite{bettencourt2007, bettencourt2010a, bettencourt2010b}.

The optimal productivity of large cities raises the question of the consequences
of urban growth to environmental quality. Indeed, it is intensely debated
whether large cities can be considered environmentally ``green'', implying that
their productivity is associated with lower than expected greenhouse gases (GHG)
and pollutant emissions \cite{bento2006, kahn2006, brownstone2009, dodman2009,
puga2010, glaeser2010a}. For instance, some of these studies report that the
level of commuting has a major contributing to the relation between GHG
emissions and city size \cite{bento2006, kahn2006, brownstone2009,
glaeser2010a}. As a consequence, compact cities would be more green due to the
attenuation of the average commuting length. More recently, however, Gaign\'e
\emph{et al.} \cite{gaigne2012} suggested that compact cities might not be as
environmentally friendly as it was thought, mainly because increasing-density
policies obligate firms and households to change place. This relocation of the
urban system then generates a higher level of pollution. In this context, here
we study the allometric laws associated with a particular type of GHG emissions
from human activity by studying the relation between CO$_2$ emissions of cities
as a function of population size. We employ a bottom-up approach combining two
unprecedented microscopic data on population and carbon dioxide emissions in the
continental US. We first define the boundaries of cities using the City
Clustering Algorithm (CCA) \cite{makse1995,makse1998, rozenfeld2008, giesen2010,
rozenfeld2011, duranton2012, gallos2012, duranton2013, ioannides2009-2013} which are
then used to calculate the CO$_2$ emissions. We find a superlinear allometric
scaling law between emissions and city size. We also explore different sectors
and activities of the economy finding superlinear behavior in most of the
sectors. Our results pertain only emissions of CO$_2$. It will be desirable to
extend it to the rest of GHGs. These results indicate that large cities may not
provide as many environmental advantages as previously thought \cite{kahn2006,
dodman2009, puga2010, glaeser2010a}.

\section*{Results}

\begin{figure}[htb]
\includegraphics*[width=0.5\textwidth]{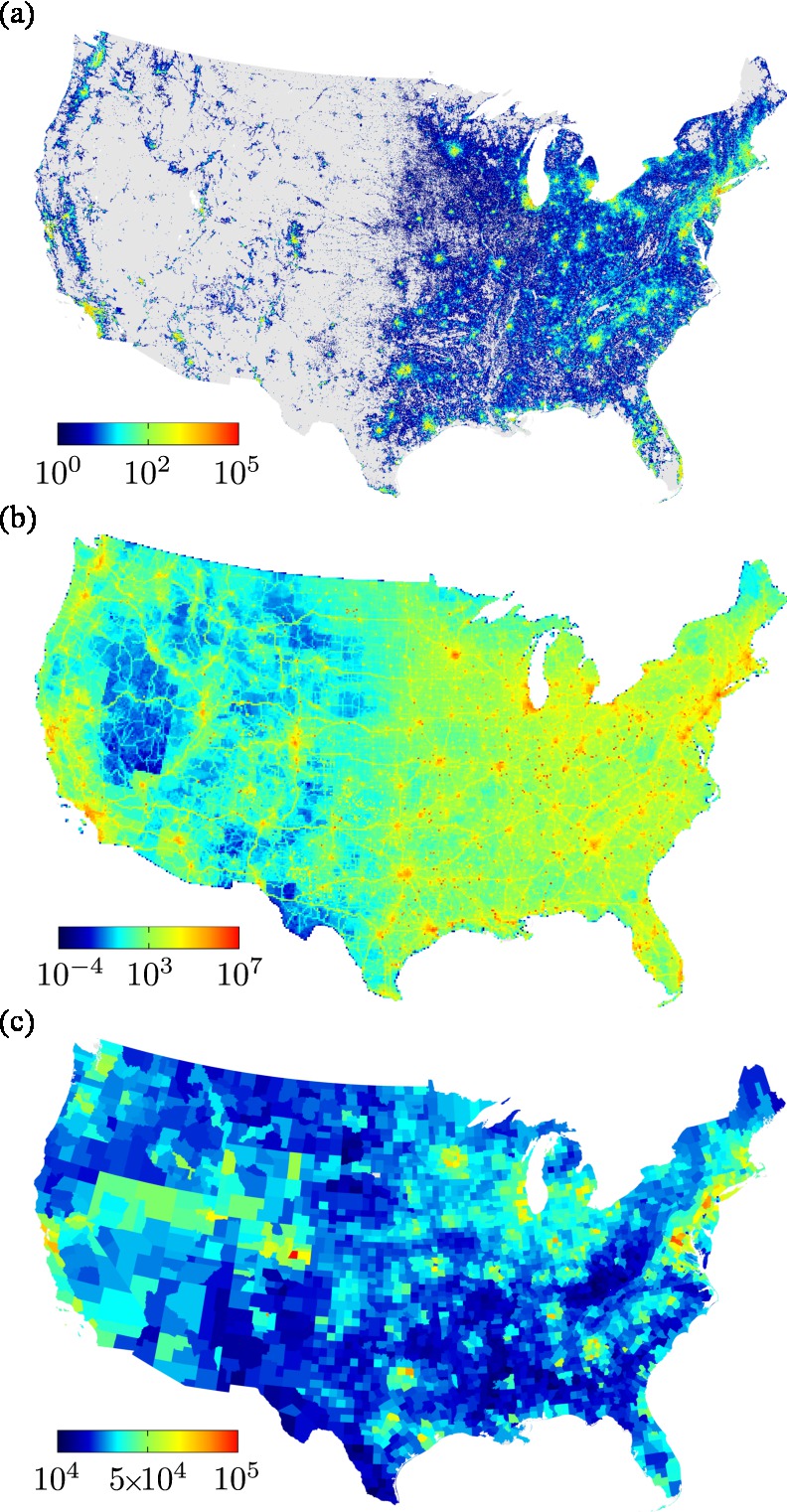}
\caption{Population and emissions in US. (a) The population map of the
contiguous US from the Global Rural-Urban Mapping Project (GRUMPv1)
\cite{grump2011} dataset in logarithmic scale. (b) The CO$_2$ emissions map of
the contiguous US from the Vulcan Project (VP) dataset \cite{vp2013} measured in
log base $10$ scale of metric tonnes of carbon per year. (c) Map of the mean
household income per capita of $3,092$ US counties in dollars from US Census Bureau dataset
\cite{censusbureau2013a} for the year $2000$.}
\label{fig1}
\end{figure}

{\bf Datasets.} We use two geo-referenced datasets on population and CO$_2$
emissions in the continental US defined in a fine geometrical grid. The
population dataset is obtained from the {\it Global Rural-Urban Mapping Project}
(GRUMPv1) \cite{grump2011}. These data are a combination of gridded census and
satellite data for population of urban and rural areas in the United States in
year $2000$ (Fig. \ref{fig1}a and Sec. 3). The GRUMPv1 data provides a
high resolution gridded population data at $30$ arc-second, equivalent to a grid
of $0.926\;km \times 0.926\;km$ at the Equator line.

The emissions dataset is obtained from the {\it Vulcan Project} (VP) compiled at
Arizona State University \cite{vp2013}. The VP provides fossil fuel CO$_2$
emissions in the continental US at a spatial resolution of $10\;km \times
10\;km$ ($0.1\;deg \times 0.1\;deg$ grid) from $1999$ to $2008$. The data are
separated according to economic sectors and activities (see Sec. 3 for
details): Commercial, Industrial, and Residential sectors (obtained from
country-level aggregation of non-geocoded sources and non-electricity producing
sources from geocoded location), Electricity Production (geolocated sources
associated with the production of electricity such as thermal power stations),
Onroad Vehicles (mobile transport using designated roadways such as automobiles,
buses, and motorcycles), Nonroad Vehicles (mobile surface sources that do not
travel on roadways such as boats, trains, snowmobiles), Aircraft (Airports,
geolocated sources associated with taxi, takeoff, and landing cycles associated
with air travel, and Aircraft, gridded sources associated with the airborne
component of air travel), and Cement Industry.

\begin{figure}[htb]
\includegraphics*[width=\textwidth]{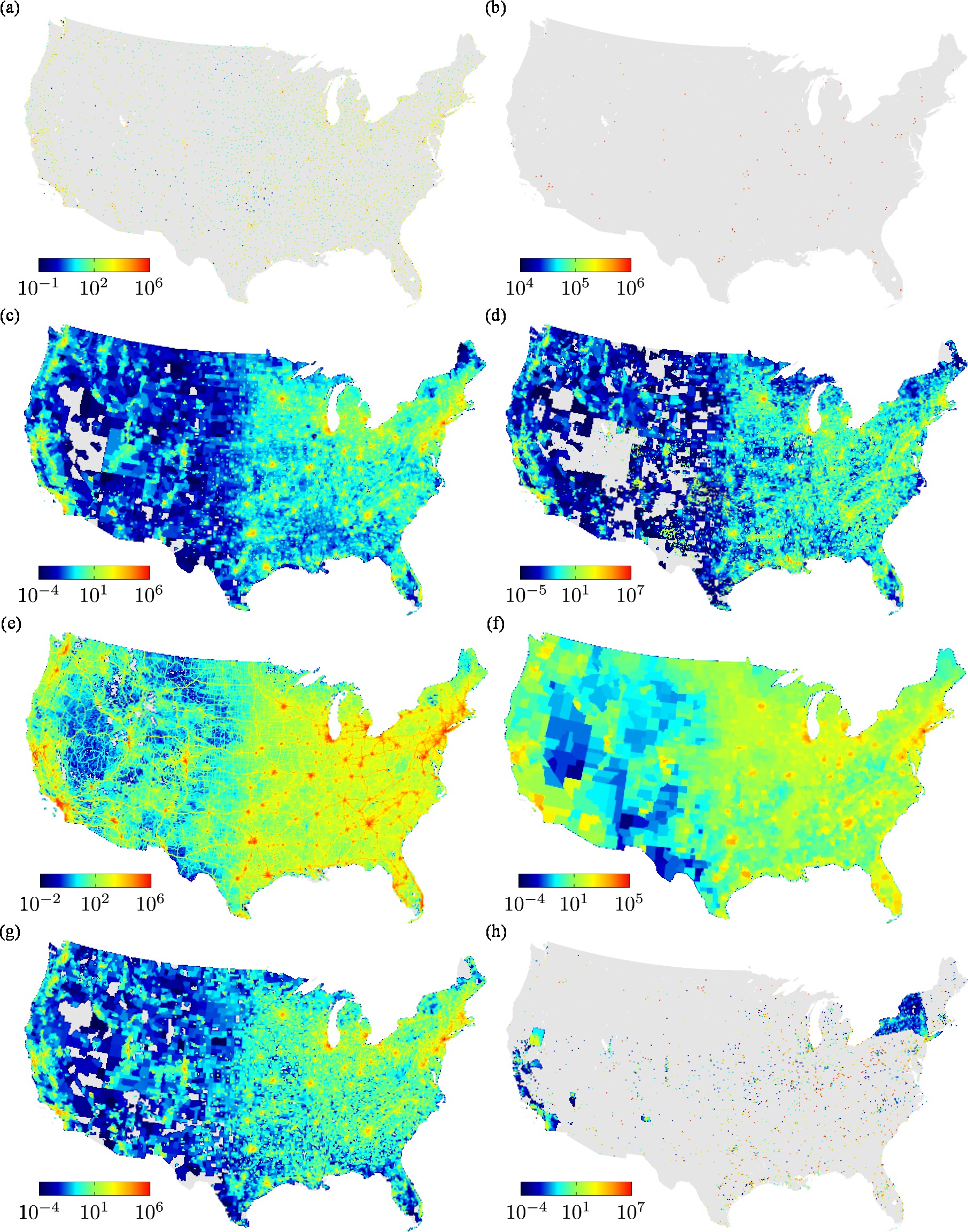}
\caption{The CO$_2$ emissions maps in metric tonnes of carbon per year from
Vulcan Project (VP) dataset \cite{vp2013} for each sector: (a) Aircraft, (b)
Cement, (c) Commercial, (d) Industrial, (e) On-road, (f) Non-road, (g)
Residential and (h) Electricity.}
\label{fig2}
\end{figure}

We analyze the annual average of emissions in $2002$ for the total of all
sectors combined (see Fig. \ref{fig1}b) and each sector separately (Fig.
\ref{fig2}). The choice of $2002$ data (rather than $2000$ as in population)
reflects the constraint that it is the only year for which the quantification of
CO$_2$ emissions has been achieved at the scale of individual factories,
powerplants, roadways and neighborhoods and on an hourly basis \cite{vp2013}.

To define the boundary of cities, we use the notion of spatial continuity by
aggregating settlements that are close to each other into cities
\cite{rozenfeld2008, giesen2010, rozenfeld2011, duranton2012, duranton2013,
ioannides2009-2013}. Such a procedure, called the City Clustering Algorithm (CCA),
considers cities as constituted of contiguous commercial and residential areas
for which we know also the emissions of CO$_2$ from the Vulcan Project dataset.
By using two microscopically defined datasets, we are able to match precisely
the population of each agglomeration to its rate of CO$_2$ emissions by
constructing the urban agglomerations from the bottom up without resorting to
predefined administrative boundaries.

We also use the US income dataset available in ASCII format by US Census Bureau
\cite{censusbureau2013a} for the year $2000$. This dataset provides the mean
household income per capita for the $3,092$ US counties. For each county, we
combined the income data and the administrative boundaries
\cite{censusbureau2013b} in order to relate them with the geolocated datasets
(Fig. \ref{fig1}c and Sec. 3).

\begin{figure}[htb]
\includegraphics*[width=0.7\textwidth]{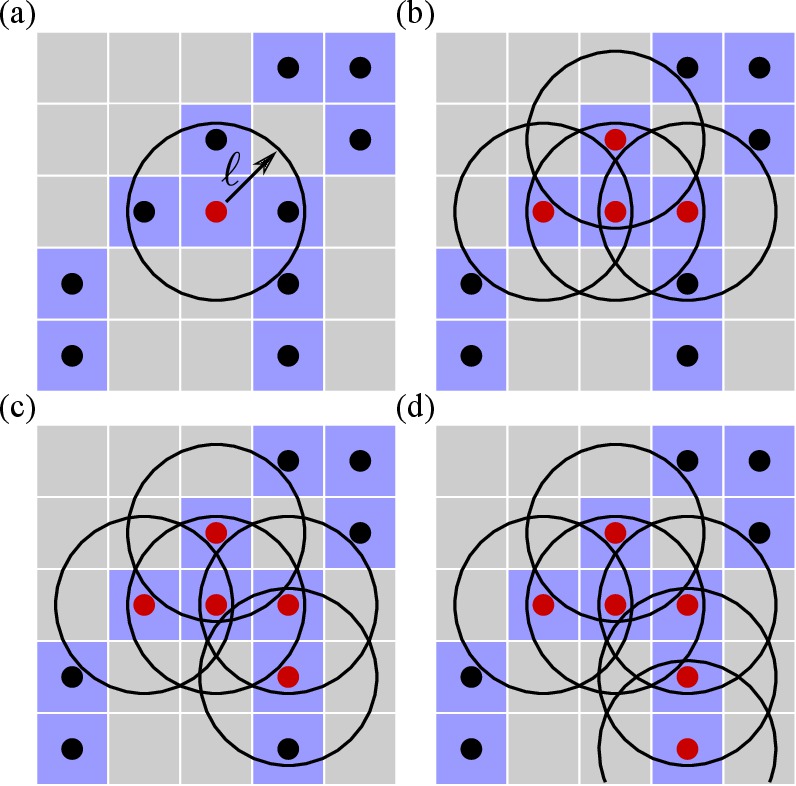}
\caption{CCA stages: We consider that if $D_i > D^*$, then the site $i$ is
populated (light blue squares). Each site is defined by its geometric center
(black circles) and the length $\ell$ represents a cutoff on the distance to
define the nearest neighbor sites. We aggregate all nearest-neighbor sites, {\it
i.e.} a CCA is defined by populated sites within a distance smaller than $\ell$
(red circles).}
\label{fig3}
\end{figure}

We first apply the CCA to construct cities aggregating population sites $D_i$ at
site $i$. The procedure depends on a population threshold $D^*$ and a distance
threshold $\ell$. If $D_i > D^*$, the site $i$ is populated. The length $\ell$
represents a cutoff distance between the sites to consider them as spatially
contiguous, i.e. we aggregate all nearest-neighbor sites which are at distances
smaller than $\ell$. Thus a CCA cluster or city is defined by populated sites
within a distance smaller than $\ell$ as seen schematically in Fig. \ref{fig3}.
Starting from an arbitrary seed, we add all populated neighbors at distances to
the cluster smaller than $\ell$ until no more sites can be added to the cluster.
The scaling laws produced by the CCA depend weakly on $D^*$ and $\ell$. and we
are interested in a region of the parameters where the scaling laws are
independent of these parameters.

This aggregation criterion based on the geographical continuity of development
was shown to provide strong evidence of Zipf's law in the US and UK
\cite{rozenfeld2008, giesen2010, rozenfeld2011, duranton2012, duranton2013,
ioannides2009-2013} in agreement with established results in urban sciences
\cite{gabaix1999, ioannides2003,giesen2011,giesen2012}. For cut-off lengths
above $\ell = 5\;km$, it was shown that CCA clusters verify the Zipf's law and
the Zipf's exponent is independent of $\ell$. Next, we first present results for
aggregated clusters at $\ell = 5\;km$, and then show the robustness of the
scaling laws over a larger range of parameter space.

In order to assign the total CO$_2$ emissions to a given CCA cluster, we
superimpose the obtained cluster to the CO$_2$ emissions dataset. If a populated
site composing a CCA cluster falls inside a CO$_2$ site, we assign to the
populated site the corresponding CO$_2$ emissions proportional to its area
$0.926^2\;km^2$, considering that the emissions density is constant across the
CO$_2$ site of $10^2\;km^2$. For a given CCA cluster, we then calculate the
population (POP) and CO$_2$ emissions by adding the values of the constitutive
sites of the cluster.

\begin{figure}[htb]
\includegraphics*[width=\textwidth]{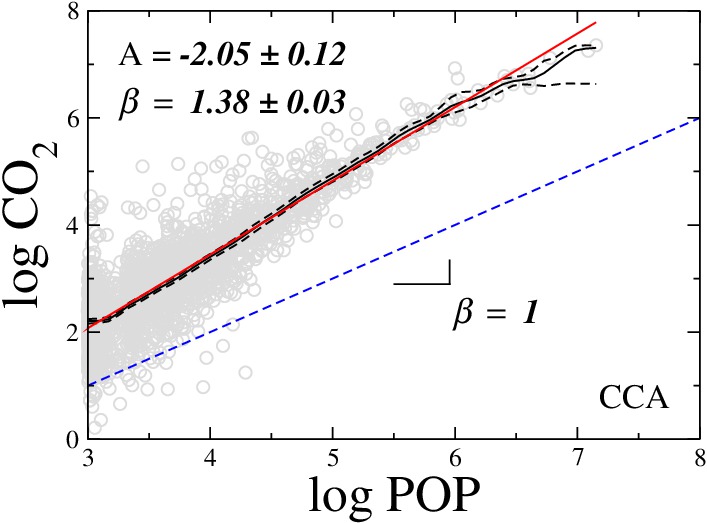}
\caption{Scaling of CO$_2$ emissions versus population. We found a superlinear
relation between CO$_2$ (metric tonnes/year) and POP with the allometric scaling
exponent $\beta = 1.38 \pm 0.03$ ($R^2 = 0.76$) for the case $\ell = 5\;km$,
$D^* = 1000$. The solid (black) line is the Nadaraya-Watson estimator, the
dashed (black) lines are the lower and upper confidence interval, and the solid
(red) line is the linear regression.}
\label{fig4}
\end{figure}

{\bf Scaling of emissions with city size.} Figure \ref{fig4} shows the
correlation between the total annual CO$_2$ emissions and POP for each CCA
cluster for $\ell = 5\;km$ and $D^* = 1000$ ($N = 2281$). We perform a
non-parametric regression with bootstrapped $95\;\%$ confidence bands
\cite{silverman1986, hardle1990} (see Sec. 3). We find that the
emissions grow with the size of the cities, on average, faster than the expected
linear behavior. The result can be approximated over many orders of magnitudes
by a power-law yielding the following allometric scaling law:

\begin{equation}\label{eq1}
\log(\rm{CO_2})=\rm{A}+\beta \log(\rm{POP}),
\end{equation}

\noindent
where $\rm{A} = 2.05 \pm 0.12$ and $\beta = 1.38 \pm 0.03$ ($R^2 = 0.76$) is the
allometric scaling exponent obtained from Ordinary Least Squares (OLS) analysis
\cite{montgomery2006} for this particular set of parameters $\ell = 5\;km$ and
$D^* = 1000$ (see Sec. 3 for details on OLS and on the estimation of
the exponent error, all emissions are measured in log base $10$ of metric tonnes
of carbon per year).

\begin{figure}[htb]
\includegraphics*[width=\textwidth]{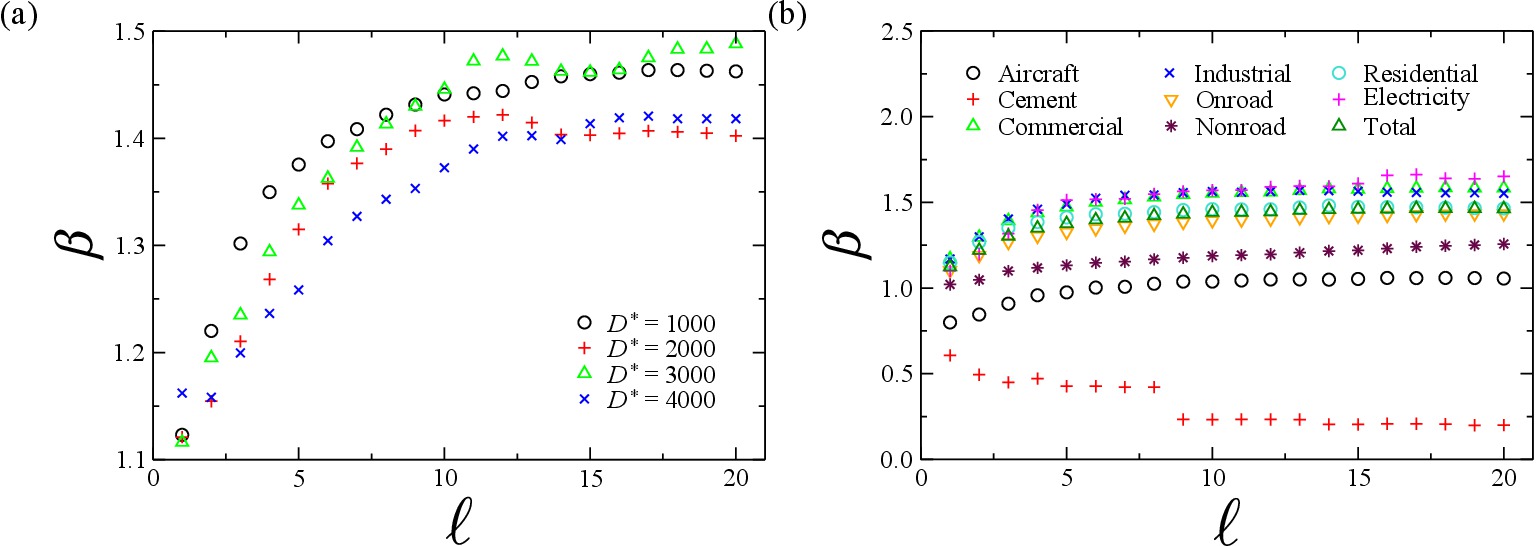}
\caption{Behavior of allometric exponent $\beta$. (a) We plot $\beta$ for the
total emissions for different $D^*$ as a function of $\ell$. The exponent
$\beta$ increases with $\ell$ until a saturation value. (b) Allometric exponent
versus $\ell$ for the different sectors of the economy as indicated. The scaling
exponent ranges from sublinear behavior ($\beta < 1$, optimal) on the cement and
aircraft sectors, to superlinear behavior $(\beta > 1$, suboptimal) on nonroad
and onroad vehicles, and residential emissions, up to the less efficient sectors
in commercial, industrial and electricity production activities.}
\label{fig5}
\end{figure}

In addition, we investigate the robustness of the allometric exponent as a
function of the thresholds $D^*$ and $\ell$. Figure \ref{fig5}a shows $\beta$ as
a function of the cut-off length $\ell$ for different values of population
threshold $D^*$ ($1000$, $2000$, $3000$, and $4000$). We observe that $\beta$
increases with $\ell$ until a saturation value which is relatively independent
of $D^*$. Performing an average of the exponent in the plateau region with $\ell
> 10\;km$ over $D^*$, we obtain $\bar{\beta} = 1.46 \pm 0.02$. Thus, we find
superlinear allometry indicating an inefficient emissions law for cities:
doubling the city population results in an average increment of $146\;\%$ in
CO$_2$ emissions, rather than the expected isometric $100\;\%$. This positive
non-extensivity suggests that the high productivity found in larger cities
\cite{bettencourt2007, bettencourt2010a} is done at the expense of a
disproportionally larger amount of emissions compared to small cities.

\begin{figure}[htb]
\includegraphics*[width=0.7\textwidth]{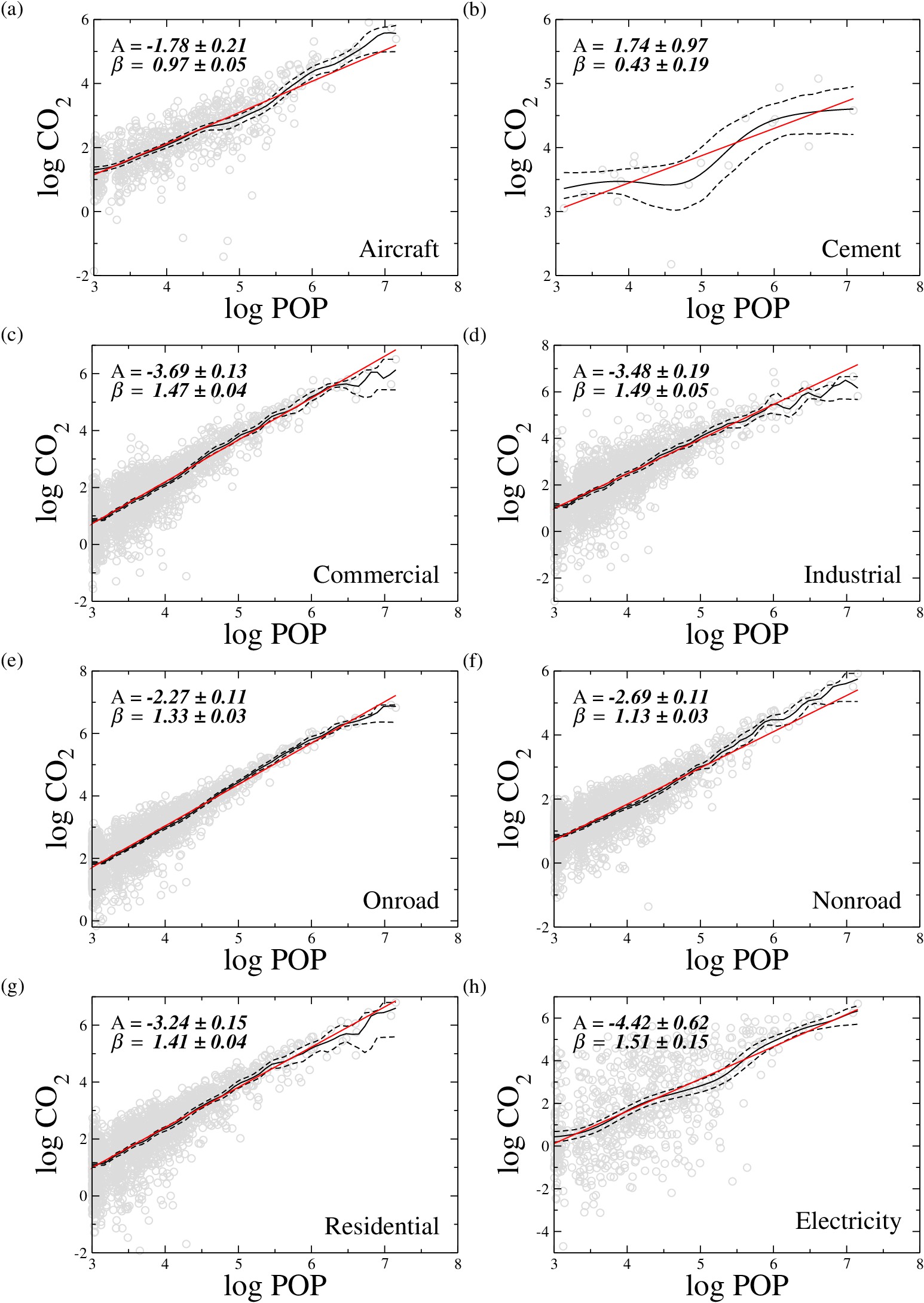}
\caption{The plot shows the CO$_2$ behavior measured in metric tonnes of carbon
per year versus POP of the CCA clusters for different sectors. We found a
superlinear relation between CO$_2$ and POP for all the cases, except to
Aircraft and Cement sectors. The solid (black) line is the Nadaraya-Watson
estimator, the dashed (black) lines are the lower and upper confidence interval,
and the solid (red) line is the linear regression.}
\label{fig6}
\end{figure}

Figure \ref{fig5}b investigates the emissions of cities as deconstructed by
different sectors and activities of the economy. We perform non-parametric
regression with bootstrapped $95\;\%$ confidence bands of $\beta$ (see Fig.
\ref{fig6} for $D^*=1000$ and $\ell = 5\;km$ by each sector) versus $\ell$ and
we find that the exponents for different sectors saturate to an approximate
constant value for $\ell > 10\;km$. We assign an average exponent, $\bar{\beta}$
over the plateau per sector as seen in Table \ref{tab1}. The sectors with higher
exponents (less efficient) are Residential, Industrial, Commercial and Electric
Production with $\bar{\beta} \approx 1.47-1.62$, above the average for the total
emissions. Onroad vehicles contribute with a superlinear exponent $\bar {\beta}
= 1.42 \pm 0.03$, yet, below the total average. The exponent for Nonroad
vehicles is also below the average at $\bar{\beta} = 1.23 \pm 0.05$, while
Aircraft sector displays approximate isometric scaling with $\bar{\beta} = 1.05
\pm 0.01$. Cement Production displays sublinear scaling $\bar{\beta} = 0.21 \pm
0.03$, although the reported data is less significant than the rest with only
$20$ datapoints of cities available.

\begin{figure}[htb]
\includegraphics*[width=0.7\textwidth]{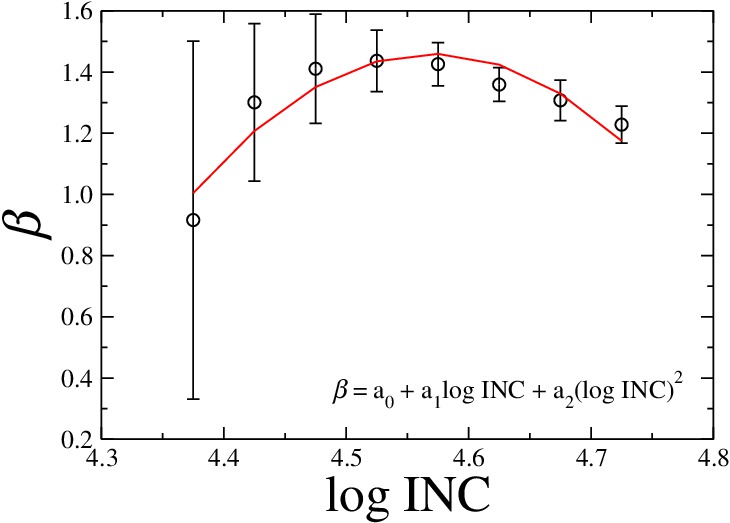}
\caption{Dependence of allometric exponent $\beta$ on the income per capita of
the CCA clusters. We found an inverted-U-shaped curve similar to an
environmental Kuznets curve (EKC). In other words, we find a decrease of the
allometric exponent $\beta$ for the lower and higher income levels, with the
following regression coefficients $\rm{a}_{\rm{0}} = -247.35$, $\rm{a}_{\rm{1}}
= 108.88$ and $\rm{a}_{\rm{2}} = -11.91$. The income turning point is located at
$10^{-\rm{a}_{\rm{1}}/(2\rm{a}_{\rm{2}})} = US\$\,37,235$.}
\label{fig7}
\end{figure}

\begin{figure}[htb]
\includegraphics*[width=0.7\textwidth]{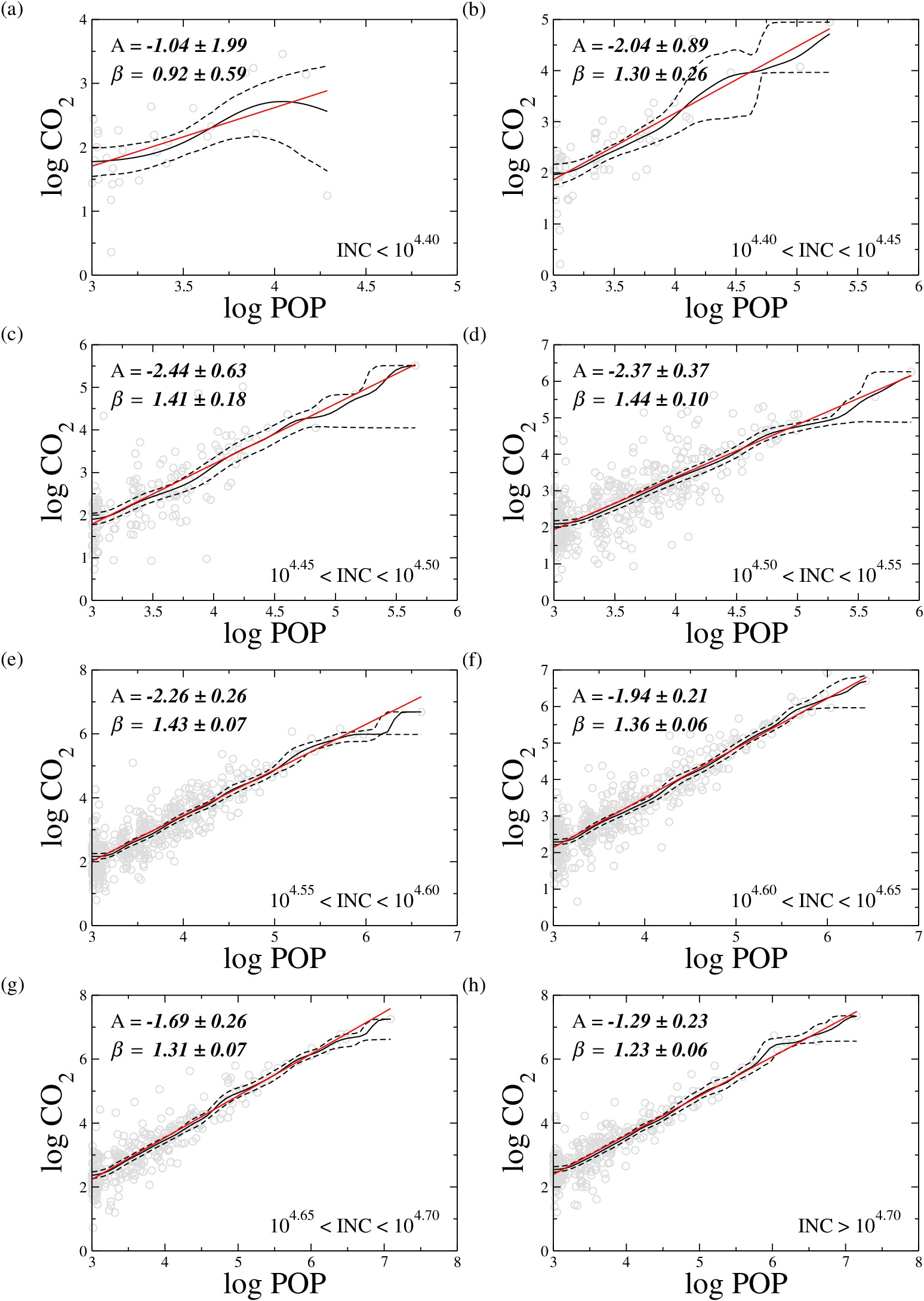}
\caption{Total CO$_2$ emissions in metric tonnes of carbon per year versus POP
of CCA clusters for different income's range as indicated. We found a
superlinear relation between CO$_2$ and POP for all the cases except for the
lowest income below \$ $25,119$. The solid (black) line is the Nadaraya-Watson
estimator, the dashed (black) lines are the lower and upper confidence interval,
and the solid (red) line is the linear regression. The resulting exponent
$\beta(\rm{INC})$ is plotted in Fig. \ref{fig7}.}
\label{fig8}
\end{figure}

We further investigate the dependence of the allometric exponent $\beta$ on the
income per capita of cities by aggregating the CCA clusters by their income
(INC) and plotting the obtained $\beta(\rm{INC})$ in Fig. \ref{fig7} (see also
Fig. \ref{fig8}). We find an inverted U-shape relationship, which is analogous
to the so-called environmental Kuznets curve (EKC) \cite{kuznets1955,
grossman1995, kahn2006}. We observe that $\beta$ initially increases for cities
with low income per capita until an income turning point located at \$ $37,235$
per capita (in $2000$ US dollar). After the turning point, $\beta$ decreases
indicating an environmental improvement for large-income cities. However, the
allometric exponent remains always larger than one regardless of the income
level (except for the lowest income) indicating that almost all large cities are
less efficient than small ones, no matter their income.

{\bf Comparison with MSA.} A further important issue in the scaling of cities is
the dependence on the way they are defined \cite{krugman1996, rozenfeld2008,
giesen2010, rozenfeld2011, duranton2012, duranton2013, ioannides2009-2013}. Thus, it
is of interest to compare our results with definitions based on administrative
boundaries such as the commonly used Metropolitan Statistical Areas (MSA)
\cite{fragkias2013} provided by the US Census Bureau \cite{censusbureau2013c}.
MSAs are constructed from administrative boundaries aggregating neighboring
counties which are related socioeconomically via, for instance, large commuting
patterns. A drawback is that MSAs are available only for a subset ($274$ cites)
of the most populated cities in the US, and therefore can represent only the
upper tail of the distribution \cite{krugman1996, rozenfeld2011, ioannides2009-2013}
(see Sec. 3 for details).

\begin{figure}[htb]
\includegraphics*[width=0.7\textwidth]{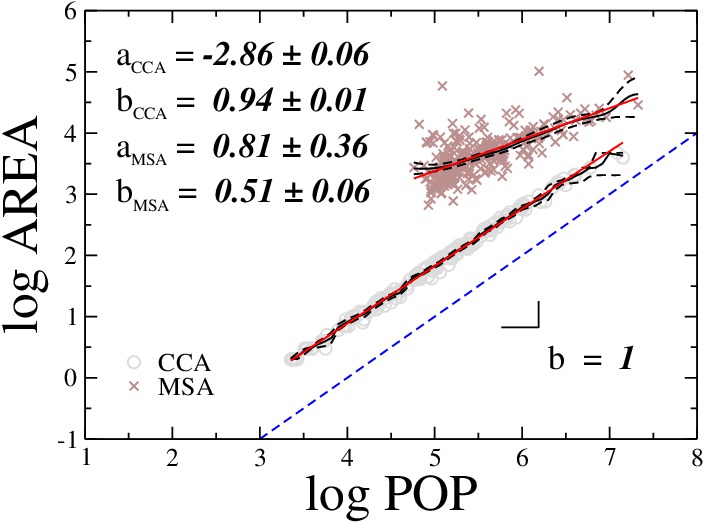}
\caption{Scaling of the occupied land area versus population for MSAs and CCA
clusters. Two problems are evident from this comparison. First, the range of
population obtained by MSA is two decades smaller than that of CCA since CCA
captures all city sizes while MSA is defined only for the top $274$ cities.
Second, the MSA violates the extensivity between land area and population while
CCA does not. This is due to the fact that MSA agglomerates together many small
cities into a single administrative boundary with a large area which can be
largely unpopulated, as can be see in the examples of Fig. \ref{fig10}. This
results in an overestimation of the size of the areas of small cities compared
with large cities, resulting in the violation of extensivity shown in the
figure. This endogenous bias is absent in the CCA definition. This bias in the
small cities ultimately affects the allometric exponent yielding a
$\beta_{\rm{MSA}}$ smaller than the one obtained using the CCAs.}
\label{fig9}
\end{figure}

Furthermore, we find that the MSA construction violates the expected extensivity
\cite{bettencourt2007, rozenfeld2011} between the land area occupied by the MSA
and their population since MSA overestimates the area of the small
agglomerations \cite{rozenfeld2011}. This is indicated in Fig. \ref{fig9}, where
we find the regression:

\begin{equation}\label{eq2}
\log(\rm{AREA}_{\rm{MSA}}) = \rm{a}_{\rm{MSA}} + \rm{b}_{\rm{MSA}}
\log(\rm{POP}_{\rm{MSA}}),
\end{equation}
with $\rm{a}_{\rm{MSA}} = 0.81 \pm 0.36$ and $\rm{b}_{\rm{MSA}} = 0.51 \pm 0.06$
($R^2 = 0.48$). This approximate square-root law implies that the density is not
constant across the MSAs: 

\begin{equation}\label{eq3}
\rho_{\rm MSA} \sim \rm{POP}^{1/2}.
\end{equation} 

On the contrary, CCA clusters capture precisely the occupied area of the
agglomeration leading to the expected extensive relation between land area and
population as seen also in Fig. \ref{fig9}:

\begin{equation}\label{eq4}
\log(\rm{AREA}_{\rm{CCA}}) = \rm{a}_{\rm{CCA}} + \rm{b}_{\rm{CCA}}
\log(\rm{POP}_{\rm{CCA}}),
\end{equation} 
with $\rm{a}_{\rm{CCA}} = -2.86 \pm 0.06$ and $\rm{b}_{\rm{CCA}} = 0.94 \pm
0.01$, with small dispersion $R^2 = 0.99$, implying that the density of
population of CCA clusters is well-defined (extensive), {\it i.e.} it is
constant across population sizes,

\begin{equation}\label{eq5}
\rho_{\rm CCA} \sim \mbox{const}.
\end{equation}
In summary, while the CCA displays almost isometric relation between population
and area, the MSA shows a sublinear scaling between these two measures. As a
consequence, the emission of CCA is independent of the population density, as
expected. On the other hand, from Eq. \ref{eq2} and Eq. \ref{eq6}, the MSA leads
to a superlinear scaling between them, ${\rm CO}_2 \sim \rho_{\rm MSA}^{1.88}$.

\begin{figure}[htb]
\includegraphics*[width=0.65\textwidth]{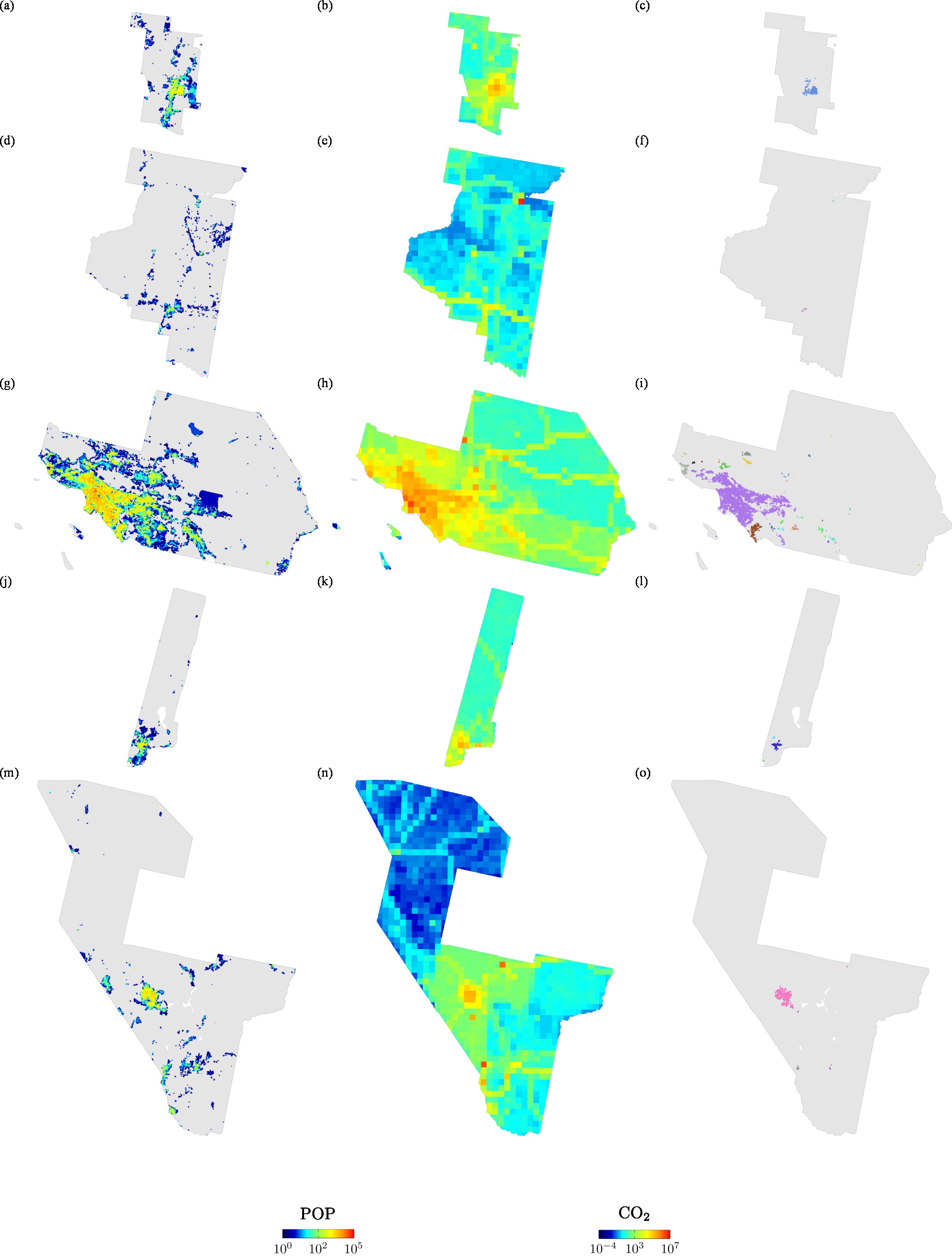}
\caption{Examples of MSA and CMSA combining the datasets from Global Rural-Urban
Mapping Project (GRUMPv1), Vulcan Project (VP) and US Census Bureau
\cite{grump2011, vp2013, censusbureau2013c}: (a)-(c) MSA of Albuquerque
(Albuquerque, NM); (d)-(f) MSA of Flagstaff (Flagstaff, AZ--UT); (g)-(i) CMSA of
Los Angeles (Los Angeles--Riverside--Orange County, CA); (j)-(l) MSA of Reno
(Reno, NV); and (m)-(o) MSA of Las Vegas (Las Vegas, NV--AZ). In the first
column, we plot the population as given by the GRUMPv1 dataset inside the
administrative boundary of the MSA as provided by the US Census Bureau. The grey
regions show the large unpopulated areas considered inside the MSA. The large
MSA areas thus put together different populated clusters into one large
administrative boundary. In the second column we plot the CO$_2$ emissions
dataset inside the boundary of each MSA. The population and the CO$_2$ emissions
are plotted in logarithmic scale according to the color bar at the bottom of the
plot. In the third column, we plot the CCA clusters inside the corresponding
MSA. Different from the MSA, the CCA captures the contiguous occupied area of a
city.}
\label{fig10}
\end{figure}

\begin{figure}[htb]
\includegraphics*[width=0.7\textwidth]{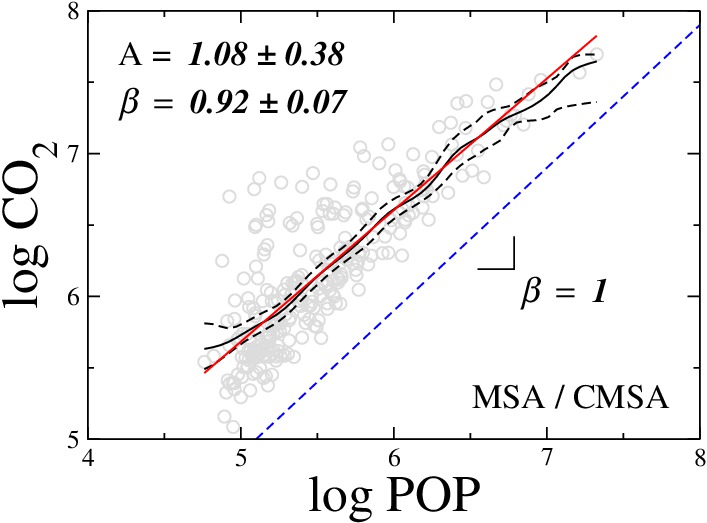}
\caption{CO$_2$ emissions in metric tonnes/year versus POP using the MSA/CMSA
definition of cities for the total CO$_2$ emissions. We found almost extensive
relation between CO$_2$ and POP with the allometric scaling exponent
$\beta_{\rm{MSA}} = 0.92 \pm 0.07$ ($R^2 = 0.71$) The solid (black) line is the
Nadaraya-Watson estimator, the dashed (black) lines are the lower and upper
confidence interval, and the solid (red) line is the linear regression.}
\label{fig11}
\end{figure}

The non-extensive character of the MSA areas is due to the fact that many MSAs
are constituted by aggregating small disconnected clusters resulting in large
unpopulated areas inside the MSA. This is exemplified in some typical MSAs
plotted in Fig. \ref{fig10}, such as Las Vegas, Albuquerque, Flagstaff and
others. The plots show that a large MSA area is associated to a series of
disconnected small counties, like it is seen, for instance, in the region near
Las Vegas. This clustering of disconnected small cities inside a MSA results
into an overestimation of the emissions associated with the Las Vegas MSA, for
instance. The same pattern is verified for many small cities, specially in the
mid-west of US, as seen in the other panels. For some large cities, like NY, the
agglomeration captures similar shapes as in the occupied areas obtained with
CCA, although it is also clearly seen that the area of the NY MSA contains many
unoccupied regions. Therefore, the occupied area of a typical MSA is
overestimated in comparison to the area that is actually populated as captured
by the CCA, the bias is larger for small cities than larger ones. This
endogeneity bias leads to an overestimation of the CO$_2$ emissions of the small
cities as compared to large cities. Consequently, we find a smaller allometric
exponent for MSA than CCA with an almost extensive relation:

\begin{equation}\label{eq6}
\log(\rm{CO_2}) = \rm{A}_{\rm{MSA}} + \beta_{\rm{MSA}} \log(\rm{POP}),
\end{equation}
with $\rm{A}_{\rm{MSA}} = 1.08 \pm 0.38$ and $\beta_{\rm{MSA}} =
0.92 \pm 0.07$ ($R^2$ = 0.71, see Fig. \ref{fig11}). This result is consistent
with previous studies of scaling emissions of MSA by Fragkias \emph{et al.}
\cite{fragkias2013}, who used MSAs and found a linear scaling between emissions
and size of the cities, and also Rybski \emph{et al.} \cite{rybski2013}, who
used administrative boundaries to define $256$ cities in $33$ countries. Table
\ref{tab2} and \ref{tab3} summarize the results of CCA and MSA cities.

Thus, the measurement bias in the MSAs leads to smaller $\beta$ found for MSA as
compared with CCA, since low-density MSAs have relatively large areas. Hence,
the CCA results, which are not subject to that endogeneity bias, should be
considered the main source of information on emissions. They show a positive
link between emissions and population size as well as the expected extensive
behavior of the occupied land. This analysis calls the attention to use the
proper definition of cities when the scaling behavior of small cities needs to
be accurately represented. Indeed, this issue arises in the controversy
regarding the distribution of city size for small cities since the distribution
of administrative cities (such as US Places) are found broadly lognormal (that
is, a power law in the tail that deviates into a log-Gaussian for small cities)
\cite{eeckhout2004, levy2009, eeckhout2009, glaeser2010b, ioannides2009-2013}, while
the distribution of geography-based agglomerations like CCA is found to be Zipf
distributed along all cities (power-law for all cities) \cite{makse1995,
makse1998, rozenfeld2008, giesen2010, rozenfeld2011, duranton2012,
duranton2013}.

\section*{Discussion}

In general, we expect that when the scaling obtained by CCA is extensive, then
any agglomeration of CCA such as MSA, should give rise to extensive scaling too.
However, when there are intrinsic long-range spatial correlations in the data
(like in non-extensive systems with $\beta\neq 1$), agglomerating populated
clusters (as done with MSA) may give different allometric exponents depending on
the particular administrative boundary used to define cities. It is of interest
to note that, beyond MSA \cite{fragkias2013}, there are other administrative
boundaries used in the literature to define cities, like for instance US-Places
studied in \cite{eeckhout2004, levy2009, eeckhout2009}. This measurement bias is
a generic property of any non-extensive system, such as a physical system at a
critical point. Thus, scaling laws obtained using administrative boundaries to
define cities which cluster data in a somehow arbitrary manner may need to be
taken with caution.

In summary, we find that CCA urban clusters in the US have sub-optimal CO$_2$
emissions as measured by a superlinear allometric exponent $\beta > 1$. The
exponent $\beta$ decreases for cities with low and high income per capita in
agreement with an EKC hypothesis \cite{kahn2006}. From the point of view of
allometry, larger cities may not represent an improvement of CO$_2$ emissions as
compared with smaller cities.

\clearpage

\section*{Methods}

\begin{figure}[htb]
\includegraphics*[width=0.7\textwidth]{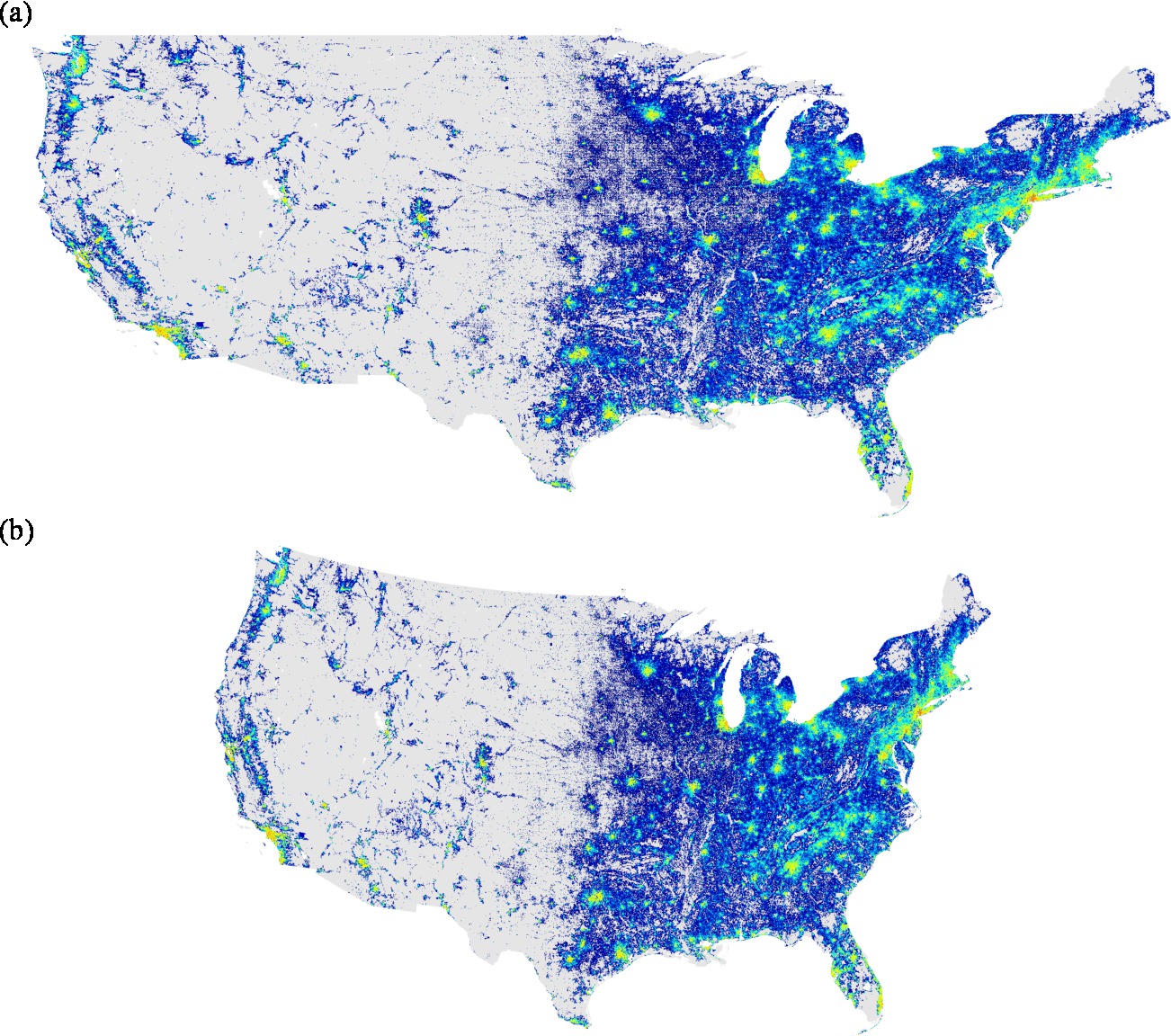}
\caption{The map projections from Global Rural-Urban Mapping Project (GRUMPv1)
\cite{grump2011}. (a) Latitude-Longitude projection and (b) Lambert Conformal
Conic projection of the population map of continental US.}
\label{fig12}
\end{figure}

{\bf Population dataset.} The United States population dataset for the year
$2000$ is a part of the {\it Global Rural-Urban Mapping Project} (GRUMPv1). The
GRUMPv1 is available in shapefile format on the Latitude-Longitude projection
(Fig. \ref{fig12}a) and it was developed by the {\it International Earth Science
Information Network} (CIESIN) in collaboration with the {\it International Food
Policy Research Institute} (IFPRI), the {\it World Bank}, and the {\it Centro
Internacional de Agricultura Tropical} (CIAT) \cite{grump2011} (Fig.
\ref{fig1}a). The GRUMPv1 combines data from administrative units and urban
areas by applying a mass-conserving algorithm named {\it Global Rural Urban
Mapping Programme} (GRUMPe) that reallocates people into urban areas, within
each administrative unit, while reflecting the United Nations (UN) national
rural-urban percentage estimates as closely as possible \cite{grump2011}. The
administrative units (more than $70,000$ units with population $>1,000$
inhabitants) are based on population census data and their administrative
boundaries. The urban areas (more than $27,500$ areas with population $>5,000$
inhabitants) are based on night-time lights data from the {\it National Oceanic
and Atmospheric Administration} (NOAA) and buffered settlement centroids (in the
cases where night lights are not sufficiently bright). In order to provide a
higher resolution gridded population data ($30$ arc-second, equivalent to a grid
of $0.926\;km \times 0.926\;km$ at the Equator line), the GRUMPv1 assumes that
the population density of the administrative units are constant and the
population of each site is proportional to the administrative unit areas located
inside of that site. We exported the original data to the ASCII format on
Lambert Conformal Conic projection (Fig. \ref{fig12}b), available to download at
\url{http://jamlab.org}. Both projections parameters are defined as follow:\\

\noindent
Projection name: Latitude--Longitude (LL)\\
Horizontal datum name: WGS84\\
Ellipsoid name: WGS84\\
Semi-major axis: $6378137$\\
Denominator of flattening ratio: $298.257224$\\

\noindent
Projection name: Lambert Conformal Conic (LCC)\\
Standard parallels: $33$, $45$\\
Central meridian: $-97$\\
Latitude of projection origin: $40$\\
False easting: $0$\\
False northing: $0$\\
Geographic coordinate system: NAD83\\

{\bf Emissions dataset.} The second dataset used in this study is the annual
mean of the United States fossil fuel carbon dioxide emissions with the grid of
$10\;km \times 10\;km$ for the year $2002$. Full documentation is available at
\url{http://vulcan.project.asu.edu/pdf/Vulcan.documentation.v2.0.online.pdf}.
This dataset was compiled by the {\it Vulcan Project} (VP) and it is already
available in binary format on the Lambert Conformal Conic projection defined
above. The VP was developed by the {\it School of Life Science} at {\it Arizona
State University} in collaboration with investigators at {\it Colorado State
University} and {\it Lawrence Berkeley National Laboratory} \cite{vp2013}. The
VP dataset is created from five primary datasets, constituting eight data types:
The National Emissions Inventory (NEI) containing the Non-road data
(county-level aggregation of mobile surface sources that do not travel on
roadways such as boats, trains, ATVs, snowmobiles, etc), the Non-point data
(county-level aggregation of non-geocoded sources), the Point data (non
electricity-producing sources identified as a specific geocoded location) and
the Airport data (geolocated sources associated with taxi, takeoff, and landing
cycles associated with air travel); The Emissions Tracking System/Continuous
Emissions Monitoring (ETS/CEM) containing the Electricity production data
(geolocated sources associated with the production of electricity); The National
Mobile Inventory Model (NMIM) containing the On-road data (county-level
aggregation of mobile road-based sources such as automobiles, buses, and
motorcycles); The Aero2k containing the Aircraft data (gridded sources
associated with the airborne component of air travel), and finally, the Portland
Cement containing the cement production data (geolocated sources associated with
cement production).

These data types supply the CO$_2$ emissions sectors: Aircraft, Cement,
Commercial, Industrial, Non-road, On-road, Residential, and Electricity. In
order to represent all the sectors in a $10\;km\times10\; km$ grid, the VP
assumes that the CO$_2$ emissions of each site is given by the contributions of
the geocoded and non-geocoded (via area-weighted proportions) sources located
inside of that site. We exported the original data to the ASCII format,
available to download at \url{http://lev.ccny.cuny.edu/~hmakse/soft_data} (Fig.
\ref{fig1}b and Fig. \ref{fig2}).

{\bf Income per capita dataset.} We also use the US income dataset available in
ASCII format by US Census Bureau \cite{censusbureau2013a} for the year $2000$.
This dataset provides the mean household income per capita for the $3,092$ US
counties. For each county, we combined the income data and the administrative
boundaries (Fig. \ref{fig1}c) in order to relate them with the geolocated
datasets. The US county boundaries are also available to download in ASCII
format by the US Census Bureau \cite{censusbureau2013b}. However, we already
joined these datasets and provided them to download at
\url{http://lev.ccny.cuny.edu/~hmakse/soft_data}.

{\bf Superimposing the datasets.} We superimposed the population and CO$_2$
datasets on the Lambert Conformal Conic projection in order to estimate the
CO$_2$ emissions on a higher grid level ($0.926\;km \times 0.926\; km$). We
checked if each population site is inside of a CO$_2$ site. If so, we assigned
the CO$_2$ value as proportional to its area ($0.926^2 km^2$), considering that
the CO$_2$ density is constant in each CO$_2$ site. For the population and
income datasets, we checked if each population site (actually, the center of
mass) is inside of some US county boundary. If so, we assigned the income value
for that site equal to the income value for the county. We performed this test
taking into account that a horizontal line (in the polygon direction), starting
in a point that is inside of a polygon, hits on it an odd number of times, while
a point that is outside of the polygon, hits on it an even number of times.

{\bf MSA.} The definitions of {\it Metropolitan Statistical Area} (MSA), {\it
Primary Metropolitan Statistical Area} (PMSA) and {\it Consolidated Metropolitan
Statistical Area} (CMSA) are provided by the US Census Bureau
\cite{censusbureau2013c}. The MSAs are geographic entities defined by some
counties socioeconomically related with population larger than $50,000$. The
PMSAs are analogous to MSAs, however they are defined by just one or two
counties also socioeconomically related with population larger than $1,000,000$.
Finally, the CMSA are large metropolitan region defined by some PMSAs close to
each other. In order to set a relation between the definition of MSA/CMSA cities
and CCA cities, we show the $15$ most populated MSA/CMSA cities and the largest
CCA cities associated to them in Table \ref{tab1} and \ref{tab2}. The largest
CCA city associated to a given MSA/CMSA is defined by the most populated CCA
city whose center of mass is inside of that MSA/CMSA boundary. All datasets are
available to download from \cite{censusbureau2013c}, including the population
and administrative boundaries of MSA/CMSA. Additionally, we make them available
at \url{http://lev.ccny.cuny.edu/~hmakse/soft_data}.

{\bf Nadaraya-Watson Method.} In order to calculate the allometric scaling
exponents, we performed well-known statistic methods \cite{hardle1990}. For one
data distribution $\{X_i, Y_i\}$, we apply the Nadaraya-Watson method
\cite{nadaraya1964,watson1964} to construct the kernel smoother function,

\begin{equation}
\hat{m}_h(x)=\frac{\sum_{i=1}^{N} K_h(x -X_i)Y_i}{\sum_{i=1}^{N} K_h(x-X_i)} \text{,}
\end{equation}
\noindent
where $N$ is the number of points and $K_h(x-X_i)$ is a Gaussian kernel of the
form,

\begin{equation}
K_h(x-X_i)=\exp{\left[\frac{(x-X_i)^2}{2 h^2}\right]}\text{,}
\end{equation}

\noindent
where the $h$ is the bandwidth estimated by least squares cross-validation
method \cite{racine2004,li2004}. We compute the $95\;\%$ ($\alpha=0.05$)
confidence interval (CI) by the so-called $\alpha/2$ quantile function over
$500$ random bootstrapping samples with replacement.

For our case, the distribution is the set of values $\{X_i,Y_i\} =
\{\log(\rm{POP}_i),\log(\rm{CO}_{2i})\}$, where $i$ is from $1$ to the number of CCA
cities $N$. Furthermore, we calculate the exponents by the ordinary least square
(OLS) method \cite{nr2007}. Let us to consider the terms,

\begin{eqnarray}
S_x=\sum_{i=1}^N X_i\text{,}\\
S_y=\sum_{i=1}^N Y_i\text{,}\\
S_{xx}=\sum_{i=1}^N X_i^2\text{,}\\
S_{xy}=\sum_{i=1}^N X_i Y_i\text{,}\\
t_i=\left(X_i-\frac{S_x}{N}\right)\text{, and}\\
S_{tt}=\sum_{i=1}^N t_i^2\text{.}
\end{eqnarray}

The regression exponents ($A$ and $\beta$ in the equation $Y = A + \beta X$) are
given by,

\begin{equation}
\beta=\frac{1}{S_{tt}}\sum_{i=1}^N t_i Y_i \quad \text{and} \quad A=\frac{S_y-\beta S_x}{N} \text{.}
\end{equation}

\noindent
If the errors are normally and independently distributed, the standard error of
each exponent is given by \cite{montgomery2006},

\begin{equation}
s.e.(A) = t_{\alpha/2,N-2} \frac{\sigma_A}{N-2}\quad \text{and} \quad s.e.(\beta) = t_{\alpha/2,N-2} \frac{\sigma_{\beta}}{N-2}\text{,}
\end{equation}

\noindent
where $t_{\alpha/2,N-2}$ is the Student-t distribution with $\alpha/2=0.025$ of
CI and $N-2$ degrees of freedom, and the variances $\sigma_A$ and
$\sigma_{\beta}$ are given by,

\begin{equation}
\sigma_A=\sqrt{\frac{1}{N}\left(1+\frac{S_x^2}{N S_{tt}}\right)}\quad \text{and} \quad \sigma_{\beta}=\sqrt{\frac{1}{S_{tt}}}\text{.}
\end{equation}

\noindent
Finally, we show the value of the regression exponents as,

\begin{equation}
A \pm s.e.(A)\quad \text{and} \quad \beta \pm s.e.(\beta)\text{.}
\end{equation}

{\bf $R$-squared.} The $R^2$ is the coefficient of determination or $R$-squared and is calculated
as following:

\begin{equation}
R^2 = 1 - \frac{\sum_{i=1}^N [Y_i - (A+\beta X_i)]^2}{\sum_{i=1}^N [Y_i - (N^{-1}\sum_{i=1}^N Y_i)]^2}
\end{equation}

\noindent
The $R^2$ by emission sector and the average $\bar{\beta}$ are in Table
\ref{tab1}.

\vspace{0.5cm}

\section*{Acknowledgements}
We gratefully acknowledge funding by NSF, CNPq, CAPES
and FUNCAP. We thank X. Gabaix and S. Alarcon for helpful discussions.

\vspace{0.5cm}

\section*{Author contributions}
EAO, JSA, and HAM designed research, performed research, and wrote the
manuscripts.

\section*{Additional information}
{\bf Competing financial interests:} The authors declare no competing financial
interests.

\clearpage

\clearpage

\clearpage

\clearpage

\clearpage

\clearpage

\clearpage

\clearpage

\clearpage

\clearpage

\clearpage

\clearpage

\clearpage
\begin{table*}[htb]
\centering
\begin{tabular}{c|c|c|c|c|c}
Sector & $ N $ & $\rm{A}$ $^\dagger$ & $\beta$ $^\dagger$ & $R^2$ $^\dagger$ & ${\bar \beta}$\\ \hline
Cement      &   20 & $ 1.74 \pm 0.97$ & $0.43 \pm 0.19$ & 0.55 & $0.21 \pm 0.03$\\
Aircraft    &  708 & $-1.78 \pm 0.21$ & $0.97 \pm 0.05$ & 0.67 & $1.05 \pm 0.01$\\
Nonroad     & 2281 & $-2.69 \pm 0.11$ & $1.13 \pm 0.03$ & 0.71 & $1.23 \pm 0.05$\\
Onroad      & 2281 & $-2.27 \pm 0.11$ & $1.33 \pm 0.03$ & 0.78 & $1.42 \pm 0.03$\\
Residential & 2280 & $-3.24 \pm 0.15$ & $1.41 \pm 0.04$ & 0.67 & $1.47 \pm 0.02$\\
Industrial  & 2276 & $-3.48 \pm 0.19$ & $1.49 \pm 0.05$ & 0.60 & $1.56 \pm 0.01$\\
Commercial  & 2281 & $-3.69 \pm 0.13$ & $1.47 \pm 0.04$ & 0.74 & $1.58 \pm 0.02$\\
Electricity &  678 & $-4.42 \pm 0.62$ & $1.51 \pm 0.15$ & 0.38 & $1.62 \pm 0.08$\\
\hline
Total       & 2281 & $-2.05 \pm 0.12$ & $1.38 \pm 0.03$ & 0.76 & $1.46 \pm 0.02$\\
\end{tabular}
\caption{Allometric exponents for CO$_2$ emissions according to different
sectors and total emissions of all sectors. We report values for $\ell = 5\;km$
and $D^* = 1000$ indicated by $^\dagger$ and the averaged value $\bar \beta$ over
$\ell > 10\;km$ and $1000\le D^* \le 4000$. The number of observed CCA clusters is
$N$.}
\label{tab1}
\end{table*}

\clearpage
\begin{sidewaystable}[htb]
\centering
\begin{tabular}{c|c|c|c|c}
CCA city & Population & Area & Income & CO$_2$ \\ \hline
New York      & 14,203,323 & 3,963 & 54,219 & 22,656,248 \\
Los Angeles   & 12,248,239 & 4,730 & 44,935 & 17,890,252 \\
Chicago       &  5,989,209 & 2,716 & 50,454 & 13,180,388 \\
San Francisco &  4,135,709 & 1,604 & 66,141 &  3,628,217 \\
Miami         &  4,041,311 & 2,029 & 38,430 &  4,851,895 \\
Washington    &  3,981,576 & 2,077 & 61,052 &  6,689,123 \\
Philadelphia  &  3,147,779 & 1,408 & 48,568 &  6,350,115 \\
Dallas        &  2,987,071 & 1,797 & 49,563 &  4,225,519 \\
Houston       &  2,670,156 & 1,520 & 43,497 &  5,104,114 \\
Detroit       &  2,534,128 & 1,578 & 47,915 &  6,038,681 \\
Phoenix       &  2,221,393 & 1,295 & 46,914 &  2,616,811 \\
Boston        &  1,838,516 &   760 & 55,055 &  3,161,289 \\
San Diego     &  1,620,953 &   744 & 48,104 &  1,881,183 \\
Denver        &  1,539,876 &   958 & 53,282 &  3,294,302 \\
Seattle       &  1,176,431 &   752 & 54,636 &  1,872,446 \\
\end{tabular}
\caption{Population ranking of the top $15$ CCA cities for $D^* = 1000$
inhabitants and $\ell = 5\;km$. The total number of cities for these parameters is
$N = 2281$. The areas are given in $km^2$, the incomes per capita are given in
US\$ and the CO$_2$ emissions are given in metric tonnes/year.}
\label{tab2}
\end{sidewaystable}

\clearpage
\begin{sidewaystable}[htb]
\centering
\begin{tabular}{c|c|c|c|c|c|c}
MSA/CMSA city & Population & Area & CO$_2$ & Population$^\dagger$ & Area$^\dagger$ & CO$_2^\dagger$ \\ \hline
New York      & 21,199,865 & 28,752 & 49,533,908 & 14,203,323 & 3,963 & 22,656,248 \\
Los Angeles   & 16,373,645 & 88,092 & 36,896,108 & 12,248,239 & 4,730 & 17,890,252 \\
Chicago       &  9,157,540 & 18,012 & 32,759,994 &  5,989,209 & 2,716 & 13,180,388 \\
Washington    &  7,608,070 & 25,304 & 26,035,616 &  3,981,576 & 2,077 &  6,689,123 \\
San Francisco &  7,039,362 & 19,462 & 15,969,389 &  4,135,709 &   207 &    379,911 \\
Philadelphia  &  6,188,463 & 15,788 & 18,462,316 &  3,147,779 & 1,408 &  6,350,115 \\
Boston        &  5,819,100 & 15,086 & 18,684,998 &  1,838,516 &   760 &  3,161,289 \\
Detroit       &  5,456,428 & 17,269 & 16,959,726 &  2,534,128 & 1,578 &  6,038,681 \\
Dallas        &  5,221,801 & 24,575 & 15,802,243 &  2,987,071 & 1,797 &  4,225,519 \\
Houston       &  4,669,571 & 21,105 & 30,483,362 &  2,670,156 & 1,520 &  5,104,114 \\
Atlanta       &  4,112,198 & 16,064 & 22,936,928 &  1,021,846 &   697 &  2,204,638 \\
Miami         &  3,876,380 &  8,748 &  6,824,965 &  4,041,311 & 2,029 &  4,851,895 \\
Seattle       &  3,554,760 & 19,834 & 10,489,945 &  1,176,431 &   752 &  1,872,446 \\
Phoenix       &  3,251,876 & 37,800 &  7,594,759 &  2,221,393 & 1,295 &  2,616,811 \\
Minneapolis   &  2,968,806 & 16,485 & 23,292,798 &  1,053,751 &   674 &  5,438,483 \\
\end{tabular}
\caption{Population ranking of the top $15$ MSA/CMSA cities and the associated
CCA ($\dagger$) for $D^* = 1000$ inhabitants and $\ell = 5\;km$. The areas are given
in $km^2$ and the CO$_2$ emissions are given in metric tonnes/year.}
\label{tab3}
\end{sidewaystable}

\end{document}